\documentstyle[aps]{revtex}
\begin{document}
\preprint{UdeM-GPP-TH-39a}
\title{Equivalence of the grand canonical partition functions of particles
with different statistics}
\author{S. Chaturvedi, R. MacKenzie$^*$, P.K. Panigrahi,  and V.Srinivasan}
\address{School of Physics, %\\
University of Hyderabad, %\\
Hyderabad - 500 046 (INDIA)}
\medskip
\address{$^*$ Laboratoire de Physique Nucl\'eaire, %\\
Universit\'e de Montr\'eal, %\\
%C.P. 6128, succ, Centre-ville\\
Montr\'eal, QC Canada H3C 3J7}
\maketitle
\begin{abstract}
It is shown that the grand partition function of an ideal Bose system 
with single particle spectrum $\epsilon_i = (2n+k+3/2)\hbar\omega$ is 
identical to that of a system of particles with single particle energy 
$\epsilon_i =(n+1/2)\hbar\omega$ and obeying a particular kind of statistics 
based on the permutation group.
\end{abstract}

%\newpage

In the recent literature, one often comes across attempts at establishing
the equivalence of dynamical systems having excitations with differing
statistics. One celebrated example is the Bose-Fermi equivalence in 1+1 
dimensional field theories$^{1}$. Another well-known case appears in 2+1 
dimensions, where an interacting theory with a Chern-Simons term of
appropriate coefficient can be made equivalent to another theory with
excitations of given spin and statistics$^{2}$. Apart from proving the
equivalence of the  partition functions, one usually constructs suitable
operators in a given theory which satisfy the canonical commutation
relations of the elementary fields of the model with which equivalence is
being sought. In this context, it is interesting to note that often an
interacting system with known statistics can be represented by a collection
of non-interacting  quasi-particles of very different statistics.  The
so-called exclusion  statistics recently advocated by Haldane falls in this
category$^{3}$. It has recently been shown that the many-body interacting
Calogero-Sutherland system (quantized as bosons or fermions) can be thought
of, in the statistical sense, as a collection of non-interacting particles
obeying Haldane's exclusion statistics$^{4}$. In many of these cases, the
composite operators of the original model with canonical commutation
relations appropriate to the  new statistics are not known. 

In the present work we show that, at the level of the grand partition 
function, an ideal Bose system with a single particle spectrum (SPS)
\begin{equation}
\epsilon_i = (2n+k+3/2)\hbar\omega~~~~,
\end{equation}
is completely identical to that of a collection
of particles with SPS
\begin{equation}
\epsilon_i = (n+1/2)\hbar\omega~~~~,
\end{equation}
but with a rather unusual
statistics based on the permutation group. Here $n$ and $k$ are positive
integers. As is clear from the SPS (1), the Bose case involves degeneracy of
the single-particle energy levels. This type of spectrum arises naturally
in elementary systems, such as
a particle in two dimensions experiencing a harmonic
potential with a frequency ratio 1:2
in the $x$ and $y$ directions respectively.
The non-relativistic single-particle Hamiltonian can be written as
\begin{equation}
H = \frac{p_x^2}{2m}+\frac{p_y^2}{2m} + \frac{1}{2}m{\omega}^2 x^2+
2m{\omega}^2 y^2~~~~.
\end{equation}
The second case (2)
can be thought of as a particle in a harmonic well in one
dimension.

We begin with a brief summary of the results obtained in Ref.~$5$, on which
the present work is based. In this work, which makes use of the ideas
developed in  Ref.~$6$, it was shown that the the general structure of the
canonical  partition function for an ideal $M$-level system corresponding
to any quantum  statistics based on the permutation group is as follows:
 
\begin{equation}
Z_N (x_1,\cdots,  x_M)  =  {\sum_{\lambda}}^{\prime} 
s_\lambda(x_1,\cdots, x_M)\,\,\,\,,
\end{equation}
where $x_i \equiv \exp(-\beta\epsilon_i),~i=1,\cdots,M$, $\epsilon_i$
denote the energies corresponding to the states $i=1,\cdots,M$, and 
$\lambda \equiv (\lambda_1,\cdots,\lambda_M)$. Here
$\lambda_1\geq\lambda_2 \geq\cdots
\lambda_M$ denotes a partition of $N$, the number of particles. 
The prime on the sum on the R.H.S. of 
$(4)$ is to indicate possible restrictions on the partitions $\lambda$ of $N$.
As we shall see later, different restrictions on $\lambda$ correspond to 
different quantum statistics. The functions $s_\lambda(x_1,\cdots,x_M)$ that 
appear in $(4)$ denote the Schur functions$^{7,8}$ which are
homogeneous, symmetric polynomials of degree $N$ in the variables 
$x_1,\cdots,x_M$. Explicitly, the Schur functions are given by 
\begin{equation}
s_\lambda(x_1,\cdots,x_M) =  {\det(x_i^{\lambda_j+M-j})\over  \det 
(x_i^{M-j})} ~~ ; ~~ 1 \le i,j \le M \,\,\,\,.
\end{equation}
(An alternative definition of the Schur functions in terms of the monomial 
symmetric functions$^{7,8}$ is

\begin{equation}
s_\chi (x_1,\cdots, x_M)= \sum_{\lambda}  K_{\chi\lambda}
m_\lambda (x_1,\cdots, x_M) \,\,\,,
\end{equation}
where $K_{\chi\lambda}$ denote the Kostka numbers$^{7,8}$.)
 
The grand canonical partition function is given by

\begin{equation}
{\cal{Z}}(x_1, \cdots, x_M, \mu ) = \sum_{N} \exp(\mu\beta N) 
Z_N (x_1, \cdots, x_M )~~~~.
\end{equation}
Using the fact that the Schur functions are homogeneous
polynomials of degree $N$, $(7)$ can be written as 
\begin{equation}
{\cal{Z}}(X_1, \cdots, X_M ;  p)= \sum_{N} 
{\sum_{\lambda}}^{\prime} s_\lambda(X_1,\cdots, X_M)~~~~~, 
\end{equation}
where $X_i \equiv \exp(-\beta(\epsilon_i - \mu))$.

The familiar quantum statistics correspond to the following restrictions on 
the $\lambda$'s

\begin{itemize}
\item[(a)] Bose statistics : $\lambda =(N,0,\cdots,0)$.

The corresponding grand canonical partition function is given by
\begin{equation}
{\cal Z}^{B}(X_1,\cdots,X_M) =\prod_i{1\over(1-X_i)}~~~~.
\end{equation}
\item[(b)] Fermi statistics: $\lambda = (1,1,\cdots,1,\cdots,0)$ 

The grand canonical partition function in this case is given by
\begin{equation}
{\cal Z}^{F}(X_1,\cdots,X_M) =\prod_i (1+X_i)~~~~.
\end{equation}

\item[(c)] Parafermi statistics$^{9}$ of order $p$: $\lambda$ such that 
$\lambda_1 \leq p $. Stated differently, $l({\lambda}^{'}) \leq p$ where
${\lambda}^{'}$ denotes the partition conjugate to $\lambda$ and 
$l({\lambda}^{'})$ denotes its length, {\it i.e.,} the number of nonzero 
${\lambda_i}^{'}$'s. ( The partition $\lambda^{'}$ conjugate to a 
partition $\lambda$ is obtained by interchanging the rows and columns of the
Young tableau corresponding to $\lambda$.)

In this case, using the results in Ref.$7$, the grand canonical partition 
function is found to be explicitly given by
\begin{equation}
{\cal{Z}}^{PF}(X_1, \cdots, X_M  ;  p) = 
{\det(X_j^{2M+p+1-i}-X_j^{i})\over  \det 
(X_j^{2M+1-i}-X_j^{i})} ~~ ; ~~ 1 \le i,j \le M ~~~.
\end{equation}

\item[(d)] Parabose statistics$^{9}$ of order $p$:  $l(\lambda) \leq p$.

The explicit expression for the grand canonical partition function 
in this case, to the best of our knowledge, is not known. 
\end{itemize}

The list given above contains only those quantum statistics based on the 
permutation group which have been extensively studied in the literature. 
For these statistics one has a second quantized formulation available with
a rather simple algebra of creation and annihilation operators. However, one
could, in principle, define new kinds of statistics, also based on the 
permutation group, by putting restrictions on $\lambda$ in $(7)$ other than 
those considered above. Some of these possibilities  are listed below:

\begin{itemize}
\item[(e)] $p$-$q$ statistics: $l(\lambda) \leq p ~~~; ~~~l({\lambda}^{'}) 
\leq q$.
 
No explicit formula for the grand canonical partition function for this case
is known.
\item[(f)] HST statistics$^{5}$: No restrictions on $\lambda$.

In this case the grand canonical partition function is explicitly given 
by$^{7}$
\begin{equation}
{\cal Z}^{HST}(X_1,\cdots,X_M) =
\prod_i{1\over(1-X_i)}\prod_{i<j} {1\over(1-X_i X_j)}\,\,\,\,.
\end{equation}

This statistics may be viewed as the $p \rightarrow \infty$ limit of parabose
and parafermi statistics.

\item[(g)] $\lambda$ even; {\it i.e.,} all $\lambda_i$'s even.

The corresponding grand canonical partition function is given by$^{7}$
\begin{equation}
{\cal Z}(X_1,\cdots,X_M) =
\prod_i{1\over(1-{X_i}^2)}\prod_{i<j} {1\over(1-X_i X_j)}\,\,\,\,.
\end{equation}

\item[(h)] $\lambda ^{\prime}$ even; {\it i.e.,}
 all ${\lambda_i} ^{\prime}$ even.

The explicit expression for the grand canonical partition function for this
case turns out to be$^{7}$
\begin{equation}
{\cal Z}(X_1,\cdots,X_M) =\prod_{i<j} {1\over(1-X_i X_j)}\,\,\,\,.
\end{equation}
\end{itemize}
 
The close similarity in the structure of the grand canonical partition
function of a Bose system (a)
and that for a system obeying the statistics defined
in (h) suggests the possibility of interpreting a system with a certain SPS 
obeying the latter statistics as a Bose system with a different SPS. This 
possibility is, in fact, realised by the SPS given in $(1)$ and $(2)$, as can
be seen from the following.

Consider a Bose system with the single-particle spectrum given by $(1)$. It 
can easily be seen that the states corresponding to the energies 
$3\hbar \omega/2$
and $5\hbar \omega/2$ are non-degenerate, the states $7\hbar \omega/2$ and 
$9\hbar \omega/2$ are doubly degenerate, the states $11\hbar \omega/2$ and
$13\hbar \omega/2$ are threefold degenerate, and so on. The
grand canonical partition function for this Bose system is therefore given by

\begin{eqnarray}
{\cal Z}^{B}&=& \left( \frac{1}{1-\alpha e^{- \beta \hbar \omega}} \right)
\left( \frac{1}{1-\alpha e^{- 2\beta \hbar \omega}} \right) 
{\left( \frac{1}{1-\alpha e^{- 3\beta \hbar \omega}} \right)}^{2}
{\left( \frac{1}{1-\alpha e^{- 4\beta \hbar \omega}} \right)}^{2} \times
\nonumber \\
&&{\left( \frac{1}{1-\alpha e^{- 5\beta \hbar \omega}} \right)}^{3}
{\left( \frac{1}{1-\alpha e^{- 6\beta \hbar \omega}} \right)}^{3}
{\left( \frac{1}{1-\alpha e^{- 7\beta \hbar \omega}} \right)}^{3} \cdots~~~.
\end{eqnarray}
where $\alpha = e^{-\beta(\hbar \omega /2 - \mu)}$.

Next, consider a system with the single particle spectrum given by $(2)$ but
obeying the statistics specified by (h). Substituting the explicit 
expressions for $\epsilon_i$ into $(14)$, one finds that the grand canonical 
partition function for this system is exactly the same as that in $(15)$
with $\alpha$ replaced by $\alpha^{'}$ where $\alpha^{'}= e^{-\beta(\hbar
\omega - 2\mu)}$. Since $\alpha$ (or $\alpha^{'}$) is to be determined by
fixing the mean number of particles, one finds that the two systems, at the 
level of grand canonical partition functions are equivalent. In other 
words, the grand canonical partition function of the first system can be 
interpreted in terms of the second system which has a different single 
particle spectrum and is governed by a different statistics. 

To conclude, we have shown the equivalence of two systems with different
single particle spectra and obeying different statistics. One of the two 
systems
has a ``complicated" single-particle spectrum (in the sense that the SPS 
has degeneracies) but obeys a ``simple" (Bose) statistics while the other
system has a ``simple" ( i.e. non-degenerate) spectrum but obeys a somewhat
exotic statistics.  
 
\vskip0.35cm
\noindent{\bf Acknowledgements:} One of us (R.M.) is grateful for hospitality 
to the School of Physics, University of Hyderabad, where this work was 
completed. 
%\newpage
\noindent{\bf References}
\begin{enumerate}
 \gdef\journal#1, #2, #3, 1#4#5#6{               % Journal reference.  Comma sets
    {\sl #1~}{\bf #2}, #3 (1#4#5#6)}            % off: name, vol, page, year
\def\pr{\journal Phys. Rev., }
\def\prb{\journal Phys. Rev. B, }
\def\prd{\journal Phys. Rev. D, }
\def\pre{\journal Phys. Rev. E, }
\def\prl{\journal Phys. Rev. Lett., }
\def\rmp{\journal Rev. Mod. Phys., }
\def\cmp{\journal Comm. Math. Phys., }
\def\mpl{\journal Mod. Phys. Lett., }

\item S.~Coleman, \prd 11, 2088, 1975; 
E.~Witten, \cmp 92, 455, 1984.
\item A.M.~Polyakov, \mpl A3, 325, 1988; P.K.~Panigrahi,
S.~Roy and W.~Scherer, \prl 61, 2827, 1988.
\item  F.D.M.~Haldane, \prl 67, 937, 1991
\item M.V.N.~Murthy and R.~Shankar, \prl 73, 3331, 1994.
\item S.~Chaturvedi \pre 54, 1378, 1996.
\item O.W.~Greenberg and A.M.L.~Messiah, \prb 138, 1155, 1965; J.B.~Hartle 
and J.R.~Taylor,  \pr 178,  2043, 1969; R.H.~Stolt and 
J.R.~Taylor, \prd 1, 2226, 1970; J.B.~Hartle, R.H.~Stolt and 
J.R.~Taylor, \prd 2, 1759, 1970.
\item I.G.~Macdonald, {\it Symmetric  functions  and  Hall  polynomials} 
      (Clarendon, Oxford, 1979).
\item  B.E.~Sagan {\it The symmetric group} (Brooks/Cole, Pacific Grove,
        California, 1991).
\item H.S.~Green, \pr 90, 270, 1953; A.M.L.~Messiah and 
O.W.~Greenberg, \prb 136, 248, 1964; S.~Doplicher, R.~Haag and 
J.E.~Roberts,  \cmp 23, 199, 1971; Y.~Ohnuki and 
S.~Kamefuchi, {\it Quantum  field theory  and parastatistics} 
(Springer-Verlag, Berlin, 1982).
\end{enumerate}
\end{document}